# Heat transport through atomic contacts

Nico Mosso[1], Ute Drechsler[1], Fabian Menges[1], Peter Nirmalraj[1], Siegfried Karg[1], Heike Riel[1] and Bernd Gotsmann[1,*]

[1]*IBM Research - Zurich, Säumerstrasse 4,8803 Rüschlikon, Switzerland*

*corresponding author: bgo@zurich.ibm.com*

**Metallic atomic junctions pose the ultimate limit to the scaling of electrical contacts[1]. They serve as model systems to probe electrical and thermal transport down to the atomic level as well as quantum effects occurring in one-dimensional systems[2]. Charge transport in atomic junctions has been studied intensively in the last two decades[2,3,4,5]. However, heat transport remains poorly characterized because of significant experimental challenges. Specifically the combination of high sensitivity to small heat fluxes and the formation of stable atomic contacts has been a major hurdle for the development of this field. Here we report on the realization of heat transfer measurements through atomic junctions and analyze the thermal conductance of single atomic gold contacts at room temperature. Simultaneous measurements of charge and heat transport reveal the proportionality of electrical and thermal conductance, quantized with the respective conductance quanta[6]. This constitutes an atomic scale verification of the well-known Wiedemann-Franz law[7]. We anticipate that our findings will be a major advance in enabling the investigation of heat transport properties in molecular junctions, with meaningful implications towards the manipulation of heat at the nanoscale.**

## Introduction

Heat transport and dissipation at the nanoscale has spurred research interest as it severely limits scaling of high performance electronic devices and circuits[8]. Atomic quantum point contacts represent an ideal platform to investigating heat transport in which quantum confinement effects cannot be neglected. The development of experimental techniques, such as scanning tunneling microscopy (STM) and mechanically controlled break junction (MCBJ) enabled the formation and manipulation of monoatomic metallic chains[3,4,5]. Using these techniques, charge transport in atomic junctions has been studied intensively[2,3,4,5], and, more recently, Joule dissipation[9,10] and thermoelectric effects[11,12] have been successfully probed. Recently, heat dissipation was measured in current-carrying single gold-gold contacts by means of STM with an integrated micro thermocouple in the tip[9]. It was shown that heat dissipates symmetrically into the two contacts, confirming that the electron transmission function *T(E)* of the junction element around the Fermi energy of the metallic contacts governs this phenomenon. Despite these recent steps forward, the properties of heat conduction through atomic junctions still remain to be fully explored.

Electrical conductance in atomic junctions is quantized. A single gold atom contact has a conductance equal to the quantum $G_0 = 2e^2/h$, where *e* is the electron charge and *h* Planck's constant. The Landauer approach used to describe charge transport can also be applied to predict heat transport[13] and predicts the validity of the Wiedemann-Franz (WF) law, which states that the thermal conductance $G_{TH}$ and the electrical conductance $G_{EL}$ are proportional to each other,

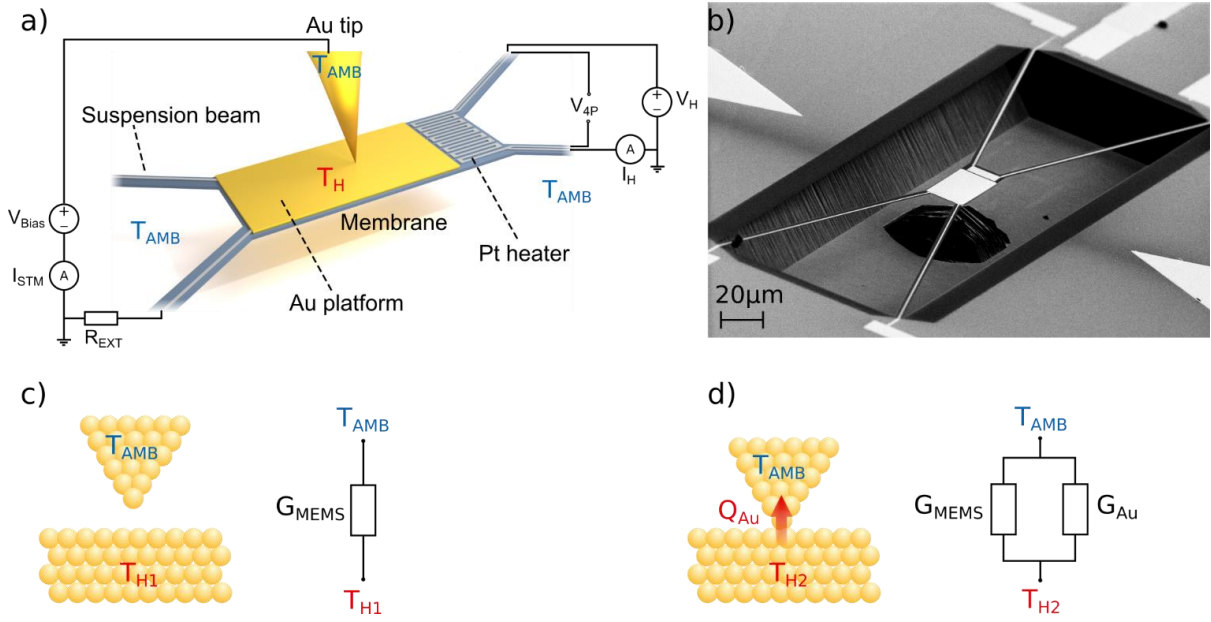

**Figure 1 | Schematic representation of the measurement technique. a)** Schematic diagram of the experiment. To monitor the temperature $T_H$ of the gold electrode, the 4-probe voltage $V_{4P}$ and the heater current $I_H$ are measured. Simultaneous measurement of the tunneling current $I_{STM}$ allows to extract the electrical resistance of the junction. **b)** SEM picture of the MEMS used in this work (See SI for exact dimensions). **c)** Prior to contact formation, the membrane is heated to $T_{H1}$. The total thermal conductance of the system is given only by the contribution of the suspension beams of the MEMS. **d)** After contact formation, the temperature of the membrane decreases to $T_{H2}$, with $T_{H2} < T_{H1}$. The total thermal conductance is now given by the sum of the thermal conductance of the MEMS ($G_{MEMS}$) and the one of the quantum point contact ($G_{Au}$).

$$G_{TH} = L_0 \cdot T \cdot G_{EL} \qquad (1)$$

where $T$ is the absolute temperature and the proportionality $L_0$, called the Lorenz number, assumes the universal Sommerfeld value $L_0 = 2.44 \times 10^{-8}$ V$^2$/K$^2$. The WF law applies to systems in which heat is predominantly transported by electrons and holds to a good approximation when scattering is mostly elastic[14]. This is largely valid for bulk metals at high and low temperature and also expected for metallic point contacts. Verification of the WF law for metallic nanostructures, however, has proved to be difficult. Both enhancements[15,16] and reductions[17,18] have been reported debating its general validity.

Marked deviations from WF law can occur when electrons are physically confined in 1D systems, in which the Fermi-liquid state is replaced by a Tomonaga-Luttinger liquid state[19]. Owing to 1D confinement of electrons, spin-charge separation occurs, enabling scattering mechanisms to affect charge and heat transport independently. It has been experimentally demonstrated recently that isolated monoatomic gold chains on surfaces host such a Tomonaga-Luttinger liquid state[20]. Although atomic junctions are well described by the Fermi-liquid theory, no experimental demonstration on the validity of WF at this scale has so far been reported. To date, most heat transport measurements in metallic ballistic systems come from 1D channels defined in 2D electron gases at low temperature[21,22]. Metallic quantum point contacts allow similar phenomena to be studied at room temperature, because of the larger energy spacing between the available conductance channels.

Investigation of the heat conduction properties at the atomic scale, however, poses big experimental challenges in terms of thermal sensitivity and mechanical stability. Here we circumvent these challenges by using a Micro Electro-Mechanical System (MEMS) with an integrated thermal sensor that is operated within a vacuum-based scanning tunneling microscope (STM) inside the IBM's Noise Free Labs[23]. The experimental technique is schematically represented in Figure 1. The system essentially combines the simultaneous measurements of heat and charge transport to extract the thermal and the electrical conductance of metallic contacts. Similarly to other STM-break junction setups[24], an STM tip is used to form and break few atom contacts on a substrate covered by a metallic layer. Here, however, the bottom electrode is suspended above a MEMS to thermally insulate it from the chip substrate. The MEMS consists of a 150-nm-thick silicon nitride membrane with four 3.5-µm-wide and 255-µm-long suspension beams. A platinum microheater is integrated on it to monitor and control the temperature of the membrane via Joule dissipation and four-point measurement of the heater electrical resistance, respectively. The temperature coefficient of resistance of the Pt film is determined independently. Next to the heater, a gold platform is fabricated as bottom electrode (see SI).

The measurements are performed under high-vacuum conditions at room-temperature, $T_{AMB}$. During the measurement, the membrane is heated to a certain temperature $T_H$ through Joule dissipation into the heater. Using a piezoelectric element, the STM tip is then approached towards the hot electrode until an electrical contact forms. As the tip is at ambient temperature ($T_{AMB}$), heat starts to flow from the hot electrode, reducing its temperature. By measuring this temperature decrease, the thermal conductance $G_{Au}$ of the contact can be extracted. The tip is then retracted until the contact is broken as described below. During the breaking process, the contact size can be reduced down to a single atom[3], as verified by simultaneous measurement of the electrical conductance quantum $G_0$. This formation and breaking process is repeated several thousands of times to obtain statistically relevant information about the quantum point contact properties.

The total thermal conductance can be calculated according to the relation:

$$G_{TOT} = \frac{\dot{Q}_{TOT}}{\Delta T} \qquad (2)$$

with $\dot{Q}_{TOT}$ representing the total heat flux provided to the membrane, which induces a temperature change $\Delta T = T_H - T_{AMB}$.

Prior to contact formation (Figure 1c), the total heat provided $\dot{Q}_{TOT}$ is equal to the effective power dissipated $P_H$ in the heater and the metallic lines and the total thermal conductance $G_{TOT} = G_{MEMS}$ corresponds to the contribution to heat conduction of the suspension beams. Once the contact has formed, the thermal conductance increases $G_{TOT} = G_{MEMS} + G_{Au}$, as shown in Figure 1d. By subtracting the in- and out-of-contact values we obtain the thermal conductance of the gold-gold contact $G_{Au}$. In calculating the total heat provided to the membrane $\dot{Q}_{TOT}$, the power dissipated in the STM circuit $P_{STM}$ has to be accounted: $\dot{Q}_{TOT} = P_H + P_{STM}$. For this, we consider symmetric power dissipation at the junction level, owing to the weakly energy-dependent transmission function of gold[9].

Clearly, the thermal conductance of the MEMS plays a crucial role in determining the sensitivity of the measurement. Indeed, $G_{MEMS}$ must be designed to be as close as possible to $G_{Au}$ so that there is sufficient contrast in the thermal conductance when only few atoms connect the tip to the membrane. However, reducing $G_{MEMS}$ means decreasing also the mechanical stiffness of the MEMS. The device used for this experiment features a thermal conductance $G_{MEMS} = 3.4 \times 10^{-8}$ W/K with an estimated vertical stiffness below 1 N/m. This stiffness is lower than the stiffness of Au-Au contacts and therefore not suitable for achieving a controlled breaking process. In fact we observe an abrupt breaking from large contacts ($G_{EL} >$ 10 $G_0$) to tunnelling. In our experiments, the tip was approached and retracted at an angle of about 25 deg with respect to the in-plane direction of the membrane. Thus, by taking advantage of the larger in-plane stiffness of the MEMS, it is possible to break the metallic contact within a displacement of the tip of a few nanometers. Note also that the gold surface exhibits some significant roughness on the order of nanometers (unlike the idealized schematic in Figure 1), which is expected to influence the breaking process. More details on the stiffness of the MEMS and the related process is provided in the Supplementary Information.

For data acquisition, we sample data during 5000 cycles of approaching and retracting the tip, measuring the electrical and the thermal conductance simultaneously. We apply a small fixed voltage bias $V_{BIAS}$ = 15-300 mV to the tip and limit the STM current with an external series resistor $R_{ext}$ = 100 kΩ. In such a configuration, the actual voltage $V_J$ at the junction depends on the junction's electrical resistance value and on the series resistance $R_{series} = R_{ext} + R_{beam}$, where $R_{beam}$ represents the electrical resistance of the metal line contacting the gold platform on the membrane. From a typical value of $R_{beam}$ ~ 30 kΩ, we obtain $V_J$ ~ 0.09 $V_{BIAS}$ at a single atom contact ($R_J$ = 12.9 kΩ).

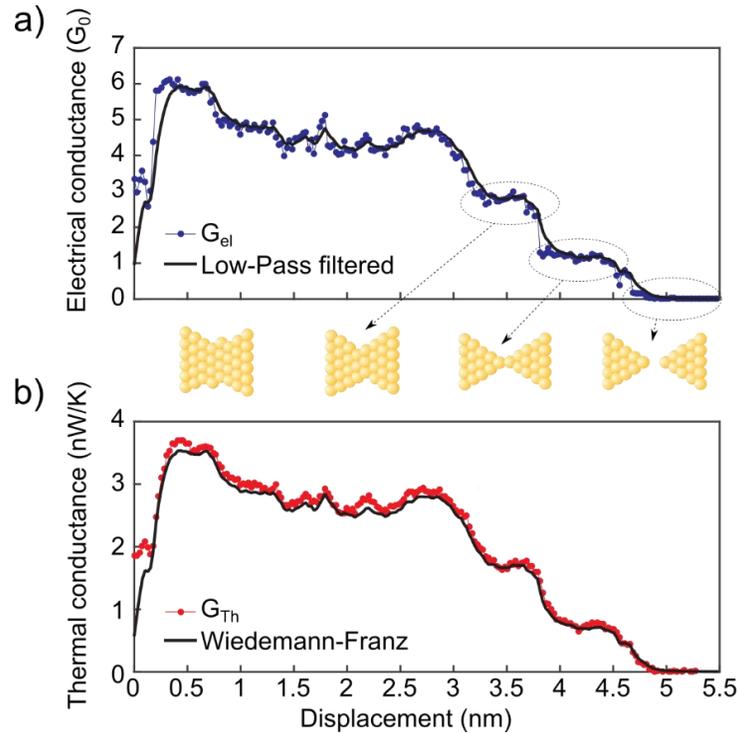

**Figure 2 | Thermal and electrical conductance of gold-gold contact during the breaking process.** The black line in the top graph, has been calculated by digitally low-pass filtering the electrical signal (blue) with the thermal time constant value $\tau$ = 30 ms from calibration. The black line in the bottom graph, has been calculated by applying the WF law to the aforementioned low-pass filtered signal. The sampling time was set to 10 ms and the tip was retracted with an angle of about 25 deg with respect to the gold surface and a speed of ~ 3nm/s. The x-axis corresponds to the displacement of the tip versus the gold surface. The total voltage applied was 50 mV, corresponding to ~5 mV at the single-atom contact owing to the series resistance. $\Delta T$ ranges from 41.5 K to 43.6 K, from closed (6 $G_0$) to open contact, respectively.

Figure 2 shows an example of an opening trace of a gold-gold junction, measured at $V_{BIAS}$ = 50 mV. In Figure 2a we observe that the initial gold-gold contact is made of few atoms featuring an electrical conductance of about 6 $G_0$. When retracting the tip, the gold contact shrinks and the electrical conductance decreases in a typical step-like fashion that is characteristic of the 1D ballistic transport regime. In particular, the plateau at around 1 $G_0$ indicates the formation of a single atom contact. After breaking the contact (at about 4.5 nm), we enter the tunneling regime. In Figure 2b the thermal conductance follows the trend of the electrical conductance trace albeit with a characteristic delay given by the thermal time constant of the MEMS of $\tau$ = 30 ms, as measured from the step response of the heater during calibration. By applying a digital low-pass filter to the electrical signal with the same τ we can directly observe that now the curves have the same time response to step features.

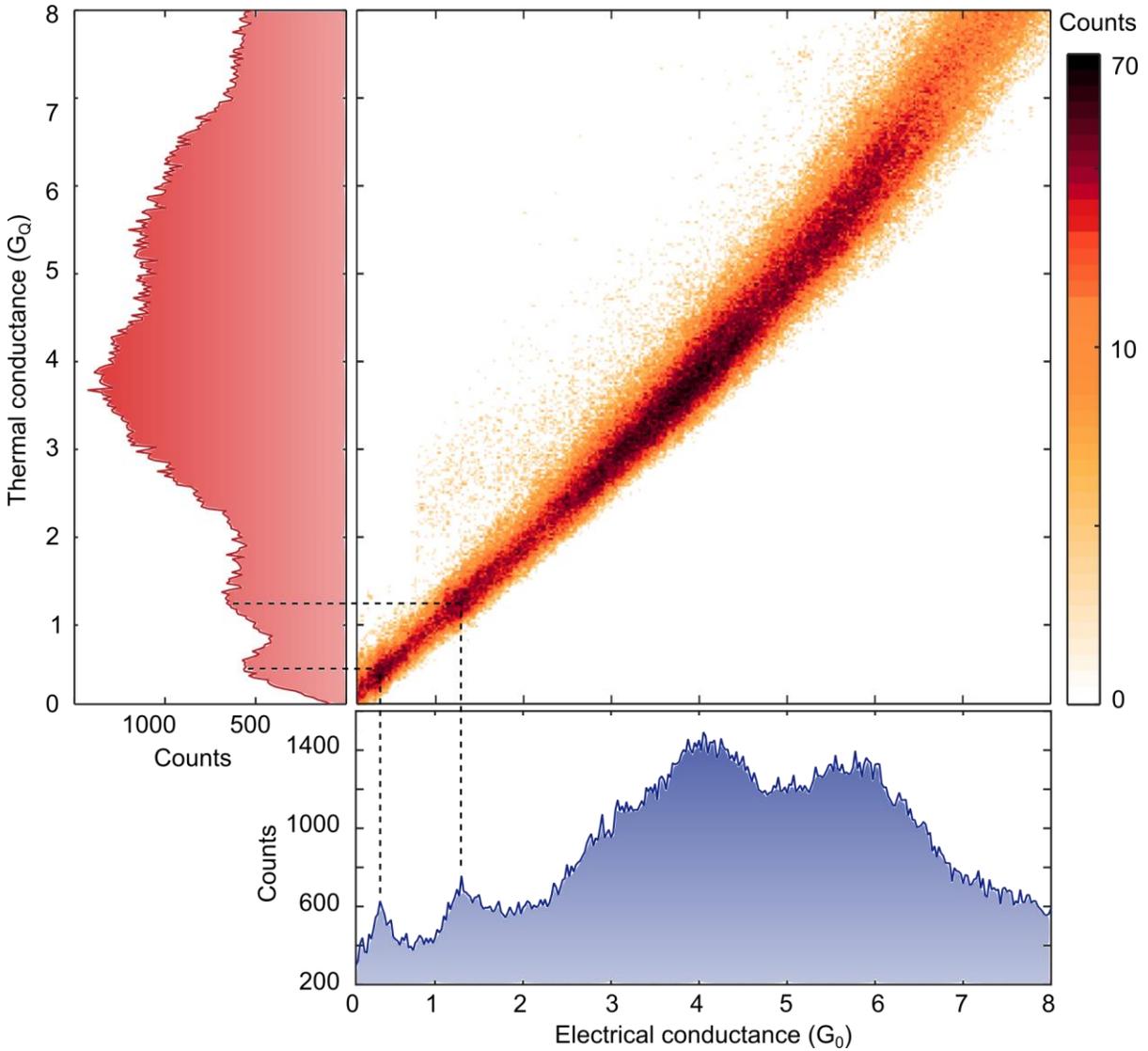

**Figure 3 | 2D histogram of thermal versus electrical conductance with respective projections onto 1D histograms.** The 2D histogram has been built from about 2000 traces selected in the temperature range ΔT = (35-50) K between tip and membrane. The thermal conductance has been normalized by the thermal conductance quantum GQ=L0TG0. The thermal and electrical 1D histograms are built with a linear binning and a bin size of 0.025 GQ and 0.025 G0, respectively.

To study the evolution of the thermal versus electrical conductance, we built two-dimensional (2D) histograms, correlating the thermal with the charge transport data as shown in Figure 3. From this graph, we can clearly deduce that the thermal and the electrical conductance are proportional to each other in such few atom contacts. The 2D histogram has been built from about 2000 traces selected from the measured 5000 ones. The selection criterion was a measured bias temperature range ΔT = (35-50) K between tip and membrane. During the 5000 traces, reconfiguration of tip and membrane surface lead to situations in which parasitic coupling of near-field radiation or molecular contamination led to "thermal short cutting" of the membrane and thereby loss of sensitivity. To verify that the analysis taking into account Joule dissipation is accurate and that the correlation between thermal and electrical transport signals is not affected by other effects than considered, we repeated the measurements using different voltage bias values and sign, reaching values of up to $V_{BIAS}$ = 300 mV. Under such high bias conditions, the contribution of the electronic dissipation at the junction is significant and must be taken into account. Furthermore, the temperature of the MEMS platform is strongly affected by the Joule effect and can even lead to a net increase of temperature in the platform instead of a net cooling through the atomic contact. The results (see SI) confirm our interpretation, and the underlying assumption that heat dissipation in atomic scale gold contacts occurs symmetrically in the electrodes, as shown recently[9]. The temperature difference across the contact varied within certain limits due to the aforementioned Joule dissipation effects and parasitic conductance effects. The temperature difference range available to us was limited to a maximum of ~50 K (to avoid buckling and mechanical instabilities of the MEMS platform). In all cases the normalization of heat current by temperature difference led to consistent values of thermal conductance.

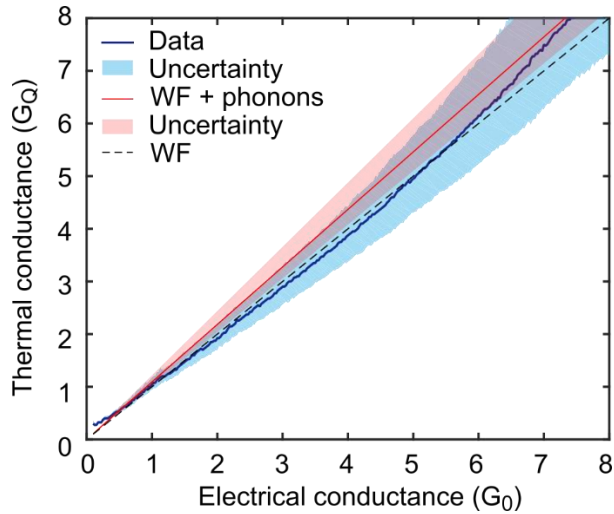

**Figure 4 | Thermal versus electrical conductance.** The thermal conductance signal has been normalized by the thermal conductance quantum $G_Q = L_0 T G_0$ where $L_0 = 2.44 \cdot 10^{-8}$ V$^2$/K$^2$ is the Lorenz number, $T = (T_H + T_{AMB})/2$ is the average contact temperature and $G_0$ is the electrical quantum of conductance. The dashed black line represents the electron contribution as predicted by the WF law, where the red continuous line takes also the phonon contribution to heat transport through the quantum point contact into account. The shaded areas represent the uncertainty regions.

In Figure 4, we plot the thermal versus the electrical conductance extracted from the median of the 2D histogram. Experimental uncertainties and systematic errors are included, taking into account the offsets of the analog-digital and digital-analog converters, the amplifier gains, Seebeck voltages and the temperature dependence of the series resistance (see SI). The measured thermal conductance comprises the contribution of all heat carriers including electrons, phonons and radiation: $G_{Au} = G_{electrons} + G_{phonons} + G_{radiation}$. The contribution of radiation between two metals[25,26] is weak, but may still reach similar magnitudes as our signal depending on the exact local geometry of the tip and surface around the junction. For blunter tips or owing to spurious contaminants, we observe a distance dependence of the thermal signal in the non-contact region (e.g. in Figure 2 for displacements greater than 5 nm). However, using etched Au tips we were able to collect data in which these effects appear only as a constant offset or are negligible (see SI for details). Phonon contribution to thermal conductance can be estimated from existing models for both single atomic contacts and larger Sharvin-type contacts, albeit with significant uncertainty. The scaling with contact size should follow approximately the scaling of electrical conductance in a situation where both charge and phonon transport is in the ballistic regime. We take into account phonon contribution between the bounds 2 and 22% of the WF law (see SI for details).

Comparing now the prediction of the WF law and the phonon contribution with our experimental data, we find quantitative agreement within the respective uncertainties. While the experimental sensitivity in principle falls within the range of the expected phonon contributions, the overall uncertainty of the total $G_{Au}$ does not allow an assessment of the possible limits of phonon transport. However, a significant violation of the WF law can be clearly excluded. Interestingly, these findings appear to be robust against variations in the contact details. For example, tunneling through small molecules may still recover the WF signature. For this, we note that the mean position of the single atom peak in Figure 3 is slightly shifted to higher values both in the electrical (~1.3 $G_0$) and thermal (~1.3 $G_Q$) conductance histograms. According to literature, the absorption of CO molecules on gold electrodes introduces an additional peak at 0.2-0.3 $G_0$ [27,28] or at 0.7 $G_0$ [29,28] depending on the experimental conditions. In particular den Boer et al.[27] reported a shift of the $G_0$ peak to larger conductance values in their study of Au-CO-Au junctions at room temperature in liquids. This shift can be attributed to the trapping of a single CO molecule in the junction for some of the measurement traces. In our case, we can expect CO molecules to be present on the Au surface even after wet chemical and heat cleaning. We can therefore interpret the increased conductance value as an electrical path through the CO molecules in parallel to the gold atomic contact. This is confirmed by the presence of a sharp peak at around 0.4 $G_0$, which can be assigned to the conduction through the molecule alone. As an important cross-check (see SI), we find that by selecting a subset of traces that exhibit plateaus in the region 0.8 to 1.2 $G_0$ (assigned to a single atom contact in the absence of a molecule) we obtain a quenched peak at 0.4 $G_0$. We note, however, that also other small molecular species such as $H_2O$, or hydrogen have been related to similar sub-$G_0$ conductance features in the literature. Therefore, further studies are needed to corroborate this preliminary conclusion on the molecular junctions.

In summary, we demonstrate the first measurements of thermal conductance of metallic quantum point contacts down to the single atomic level at room temperature. The experimental results confirm the validity of the Wiedemann-Franz law in 1D systems as predicted by the Landauer approach for heat transport. To achieve these results, we developed a unique experimental technique that allows the simultaneous measurement of heat and charge transport in atomic contacts. By further improving such technique it may be possible to investigate also phonon transport in single atomic scale contacts, which may be used to verify conclusions drawn from previous studies of mechanical contacts of atomically rough surfaces[30]. This will also enable heat transport to be investigated not only in quantum point contacts of different metals, but also in molecular junctions, a fundamental scientific and technological step forward in managing and controlling heat at the nanoscale.

# Acknowledgements

We acknowledge funding by the European Commission (EC) FP7 ITN "MOLESCO" Project No. 606728. We would like to thank Jascha Repp, Kirsten Moselund and Walter Riess for the management support on the project. We acknowledge technical support from Meinrad Tschudy, Heiko Wolf, Emanuel Lörtscher, Steffen Reidt, Antonis Olziersky, Gerhard Meyer and Charlotte Bolliger. We thank Colin Lambert, Hatef Sadeghi, Giorgio Signorello, Fabian Motzfeld and all the MOLESCO partners for fruitful discussions concerning this work.

# Heat transport through atomic contacts - Supplementary Information

Nico Mosso, Ute Drechsler, Fabian Menges, Peter Nirmalraj, Siegfried Karg, Heike Riel and Bernd Gotsmann[*]

*corresponding author: bgo@zurich.ibm.com

## Contents

## Fabrication process and sample preparation

For the fabrication of the MEMS devices, we follow the process outlined by Karg et al.[1]. We start from a silicon wafer coated with 150 nm Low-Stress $SiN_x$ (SiMat, Germany). We then thermally evaporate 25 nm of Pt with 2 nm of Cr as adhesion layer. We pattern the Pt layer by e-beam lithography to fabricate the heater and the suspended metal lines. Afterwards, we thermally evaporate 50nm of Au and, using optical lithography, we define the platform on the MEMS and the contact pads. Finally, we perform a last optical lithography step to define and release the MEMS by underetching the $SiN_x$ layer in tetramethylammonium hydroxide (25% concentration for 30 min at 80 °C) and subsequent replacement of solvent using isopropanol, which is then removed in a critical point dryer. The resulting MEMS structure consists of four $SiN_x$ beams of length 255 µm, thickness t = 150 nm and width w = 3.5 µm. The beams carry Pt lines of thickness 25 nm and width 100 nm. The platform features a heater with a Pt line of width 150 nm and 38 meanders. Furthermore, it comprises an Au platform of approximately 30 µm x 100 µm.

The tip is prepared by electrochemical etching of a 0.25 mm diameter gold wire with 99.99+% purity (GoodFellow) using the procedure reported by Boyle et al.[2]. Cleaning the gold surface on such suspended MEMS devices is a challenging task, because the usual cleaning steps ($O_2$ plasma, flame annealing) can damage them. Thus, prior to starting the measurement, we usually anneal the membranes for 2-3 h at about 200 °C in high vacuum ($\approx 10^{-7}$ mbar), to desorb organic contaminants that accumulate on the gold surface under standard laboratory conditions[3]. In this way we managed to obtain reproducible one-dimensional (1D) electrical conductance histograms showing quantization peaks, despite visible traces of small molecules (e.g. CO) as discussed below.

## Stiffness estimation of the MEMS device

To estimate the order of magnitude of the stiffness k of our MEMS structure, we model it as two SiN beams of length $l$ = 510 µm, thickness $t$ = 150 nm and width $w$ = 3.5 µm. Each beam carries two Pt lines of thickness 25 nm and width 100 nm, modelled as independent springs. The mechanical parameters used for SiN are: Young's modulus $E$ = 260 GPa[4], tensile stress $s$ ~75 MPa (estimated, ultra-low stress nitride, SiMat, Germany), and for Pt: Young's modulus 170 GPa, tensile stress ~500 MPa (typical value for evaporated Pt). During operation, additional thermal stress arises governed by the coefficients of thermal expansion of SiN of $2 \times 10^{-6}$ $K^{-1}$ and Pt of $9 \times 10^{-6}$ $K^{-1}$ [4].

Using the linear elastic, continuum theory expressions for double-clamped beams under stress[5]

$$k_{\text{beam}} = \frac{16Ewt^3}{3l^3} + \frac{swt}{l} \quad (3)$$

we obtain a stiffness in out-of-plane direction of ~0.15 N/m, which is dominated by the contribution from stress (second term in the equation). However, we note that the resulting thermal stresses can be of similar order of magnitude as the tensile film stress resulting in large uncertainties in this estimation. Nevertheless, this stiffness is clearly below the one of Au-Au contacts (~1.5 N/m)[6] and therefore not suitable for achieving a controlled breaking process. Instead it supports our observation of abrupt breaking from large contacts ($G_{EL} > 10\ G_0$) to tunnelling during retraction of the tip.

To calculate the stiffness in the in-plane direction, we note that the beams are angled with respect to the motion direction of the tip and that the four half-beams form a cross shape. Therefore contributions of both "in-plane and perpendicular to beam" ($k_{\text{perp}}$) and "in-plane and along beam" ($k_{\text{along}}$) stiffness are used. $k_{\text{perp}}$ can be calculated by using equation (3) (swapping $t$ and $w$) while $k_{\text{along}}$ can be defined as

$$k_{\text{along}} = \frac{Ewt}{l} \qquad (4)$$

These enter the total stiffness with a $\cos(\beta)$ and a $\sin(\beta)$ as multiplication factor, respectively, where beta is the angle with respect to the motion direction ($\beta = 10°$). Accordingly the in-plane direction is significantly stiffer with ~100 N/m. The large value is due to the angled arrangement of the four beam-section. In this direction a lateral displacement must lead to an elongation of the beam sections rather than merely a simple bending.

We would like to point out that the estimations given here carry significant uncertainties since we do not consider other displacement modes such as bending and torsional modes that may contribute to the effective stiffness. Nevertheless, our estimations serve to explain the empirical finding of a very different in-plane and out-of-plane stiffness resulting, for example, in different spring-loading distances in the two directions.

## On the choice of the angled approach direction

As described in the main manuscript, the approach/retraction motion was performed at an angle of 25 deg with respect to the in-plane direction of the membrane. When the tip was moved without such an angle, i.e. simply perpendicular to the surface, then most of the times the pulling motion led to strong spring loading of the sample. As a result, the conductance value typically jumped directly from the values of the fully closed junction (e.g. 10 $G_0$ or more) to the completely open junction with conductance values below $10^{-3}$ $G_0$, effectively omitting the regime of atomic-scale contact sizes.

If, in contrast, the angled approach was used, then the spring loading of the MEMS sample was significantly reduced. A remaining small amount of spring loading occurred potentially ensuring that the opened junction does not immediately close again due to long-range attractive forces through a jump-to-contact procedure. More importantly, however, the pulling procedure resulted in a higher yield in producing junctions with conductance values associated with atomic scale contacts between 1 and 10 $G_0$. The observation that a rupture, or decohesion process is eased by a lateral motion is well known from tribology studies. However, the downside of this approach is a greater sensitivity of the measurement to variations in the parasitic contacts caused by radiation and spurious contamination, as described in the main text. A significant influence of the roughness of the sputtered gold surface on the observed decohesion process is also expected. The surface roughness was measured using atomic force microscopy with a sharp tip (a nanosensors PPP-NCHR-W, guaranteed radius of curvature below 10 nm taken fresh from the wafer). We found an rms roughness of 2.5 nm measured over an area of 1x1 $\square$m². On the atomic scale this can translate into significantly different local angles with respect to an idealized flat surface, influencing the stretching of bonds during the opening traces.

The estimations of the in-plane and out-of-plane spring constants of the MEMS device support this reasoning. To quantify the effect, however, would not only require a more quantitative determination of spring constant than given in this study, it would also require knowledge on the roughness and geometry of both tip and surface that goes beyond the scope of this work.

## On the temperature of the metal tip

We assume that the gold tip is at room temperature near the apex, that is at a distance from the atomic junction consistent with the ballistic nature of transport, i.e. comparable to the mean free path of electrons on the order of 4 nm[7]. If there was a thermal resistance in series along the Au wire, i. e. between the junction and the base of our microscope, there could possibly be a temperature difference between the two positions. For an order of magnitude estimation of the effect we calculate the thermal resistance of a wire shaped as a truncated cone with a full opening angle of ~90 degrees (estimated from electron microscopy images of our tips) and use Fourier's law of (diffusive) thermal transport. Assuming a thermal conductivity of 315 W/(Km) we estimate a resistance of ~5x10$^5$ K/W. As this is orders of magnitude smaller than our measured values we may safely neglect the effect of a series thermal resistance in the tip within the accuracy of our experiment.

The temperature of the base plate of the microscope is the reference for both tip and MEMS. Therefore, any uncertainty in the knowledge enters only via the absolute temperature (~300 K) in the estimated phonon contribution while the electrical conductance quantum is independent of temperature. The room temperature of the lab is controlled to within 0.05 K [8], and the power dissipated in the setup is not sufficient for significantly elevating the temperature of the microscopes base plate, which is thermally well coupled to the environment. The uncertainty in base temperature can therefore be neglected within the uncertainty of this experiment.

The position of the G0-peak in the electrical conductance histogram should not depend on the temperature of the tip. This is not true for the thermal conductance quantum which by definition depends linearly on the junction temperature. However, the results obtained at high bias (**Fig. S5b**) confirm within the uncertainties that a large temperature change at the tip can be excluded.

## Details of the data analysis

The temperature of the heater is extracted by measuring its 4-probe resistance $R_{4P}$:

$$T_H - T_{AMB} = \frac{(R_{4P} - R_{4P0})}{\alpha \cdot R_{4P0}} \quad (3)$$

where $R_{4P0}$ corresponds to the 4-probe resistance at ambient temperature $T_{AMB}$. $R_{4P0}$ is calculated from the linear fit of the *I-V* characteristic of the device performed in vacuum as shown in the left panel of **Fig. S1**. The temperature coefficient of resistance, α, is measured on a Pt line fabricated on the same wafer as the MEMS devices from the linear fit of the resistance versus temperature $R(T)$ curve, as shown in the right panel of **Fig. S1**. Following this approach we obtain $R_{4P0} = (35.4 \pm 0.1)$ kΩ and $\alpha = (1.283 \pm 0.001) \times 10^{-3}$ K$^{-1}$ for the device under study.

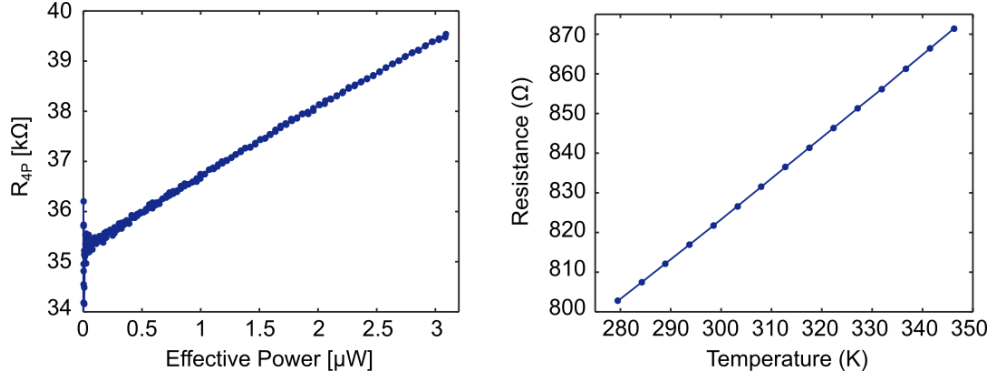

**Fig. S1| Calibration curves.** (Left) *I-V* calibration: the 4-probe heater resistance is measured at different power values in high vacuum conditions. From the linear fit we can find the resistance value at ambient temperature, $R_{4P0}$, needed to convert the resistance change into a temperature difference *ΔT*. (Right) Temperature coefficient of resistance (α) calibration: the resistance of a Pt line is measured at different *T*. From the linear fit, we can extract the value of α for the Pt film.

As stated in the main article, we measure the thermal conductance of the gold contact $G_{Au}$ using the difference between the out-of-contact ($G_{TH} = G_{MEMS}$) and in-contact ($G_{TH} = G_{MEMS} + G_{Au}$) values. The thermal conductance $G_{TH}$ is equal to the ratio between the total heat provided to the MEMS structure $\dot{Q}_{TOT}$ and the temperature difference $\Delta T$ across the junction (note that tip and substrate are at the same temperature $T_{AMB}$). We assume that $\dot{Q}_{TOT} = P_{EFF} = P_H + P_{STM}$ where the total effective power dissipated on the MEMS $P_{EFF}$ is equal to the sum of the heater power $P_H$ and STM power $P_{STM}$. These power values are calculated from the respective current measurements with the following relations

$$P_H = R_{4P} I_H^2 + \frac{1}{2} r_H (R_{2P} - R_{4P}) \cdot I_H^2 \qquad (4)$$

$$P_{STM} = \frac{1}{2}(R_J + r_{STM} R_b) \cdot I_{STM}^2 \qquad (5)$$

with $R_{4P}$, $R_{2P}$ being respectively the 4-probe and 2-probe heater resistance values, $R_J$ the junction resistance, $R_b$ the resistance of the metallic line contacting the gold pad on the membrane. $r_H$ and $r_{STM}$ are two correction factors accounting for the fact that only the portion of the metallic lines on the suspension beams contributes to the overall heat provided. From the design layout of the sample we estimated them to be 0.83± 0.04 and 0.87± 0.04, respectively.

Note that in equation (5) we assume that heat is dissipated symmetrically in the atomic junction, so that only half of the electric power contributes to the total heat. This result is valid both in the tunneling regime and in the contact regime because transport occurs elastically. Specifically, because gold has a very low energy dependent transmission function *T(E)* half of the STM power is dissipated in the membrane and half in the tip when working with low bias voltages[9].

## Measurement system

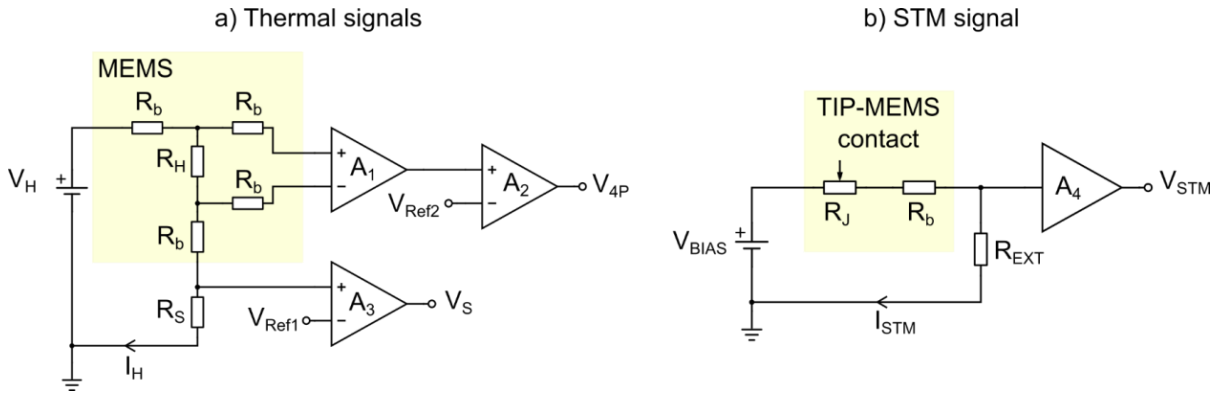

**Fig. S2 | Schematic of the circuits used to measure (a) the thermal and (b) the electrical conductance of the metallic contacts.** a) Measurement circuit of the 4-probe heater resistance, $R_H$, and the heater current, $I_H$. $R_b$ represents the resistance of the metallic line on the suspension beams, $R_S$ is an external series resistance, and $V_{Ref1}$ and $V_{Ref2}$ are two voltage signals used as reference for the differential amplifiers $A_3$ and $A_2$, respectively. $A_1$, $A_2$ and $A_3$ correspond to the Stanford Research SR640 differential filters/amplifiers. b) The STM current, $I_{STM}$, is measured from the voltage drop over the series resistance, $R_{EXT}$. $A_4$ is a Stanford Research SR640 filter/amplifier used in a single-input configuration.

**Fig. S2** shows a schematic of the measurement circuit used. For the analog inputs ($V_H$, $V_{BIAS}$, $V_{Ref1}$, $V_{Ref2}$) and outputs ($V_{4P}$, $V_S$ and $V_{STM}$) we use, respectively, the 18-bits ADC card (Analog to Digital Converter) and the 16-bit DAC card (Digital to Analog Converter) of the data acquisition system (ADwin-Pro II, Jaeger, Germany). Prior to starting a measurement the voltage offsets of all the inputs are acquired and stored. The offsets of the output signals and the external resistors are characterized separately using the Digital Multimeter (Keithley 196). To increase the measurement accuracy, the voltage gains of the filters/amplifiers ($A_{1-4}$) (Stanford Research Systems SR640) were calibrated and the correction factors stored for data analysis. Note that thanks to the high temperature stability of the IBM Noise Free Labs ($\approx 0.1$ K), thermal drifts of the electronic instruments are minimized, increasing the over-time reliability of the calibrations performed.

Modulation of thermal bias or voltage bias combined with phase-sensitive detection (lock-in amplification) is commonly done to improve sensitivity by measuring differential $G_{TH}$ or $G_{EL}$. In this experiment, however, measurements are performed with a compliant mechanical structure. Modulation of sensing (heating) current or bias voltage may lead to significant vibrational motion of the platform. The interpretation of differential signals therefore may be conflicted. Significant effort was made to reach sufficient sensitivity for DC measurements in the given bandwidth of the experiment of 30 ms.

## Evaluation of the STM current

To measure the current flowing through the junction $I_{STM}$, we need to take into account two further correction factors affecting the actual voltage at the junction. The first one is the temperature dependent value of the beam resistance $R_b(T)$ and the second one is the Seebeck voltage $V_{Seeb}(T)$ created at the Pt-Au overlap on the membrane. The junction electrical resistance is therefore calculated with

$$R_J = \frac{V_{BIAS} + V_{Seeb}(T)}{I_{STM}} - R_{ext} - R_b(T) \tag{6}$$

where $R_{ext}$ is the external series resistance used to measure and limit the STM current $I_{STM}$. By heating up the MEMS, a change in resistance is induced in all suspended metallic lines because of the temperature gradient created. This also affects the value of the series resistance in the STM circuit as $R_b(T) = r_{STM}R_{b0}\left(1 + \frac{\alpha(T-T_0)}{2}\right) + (1 - r_{STM})R_{b0}$ where $T$ is the temperature at the membrane, $T_0$ the one at the substrate, corresponding to room-temperature, and $R_{b0}$ is the beam resistance value at $T_0$. Note that only the suspended fraction $r_{STM}R_{b0}$ is affected by the temperature gradient. Thus by knowing T, we can correct for this offset. In this study, $R_{ext}$ was set to 100 kOhm and $R_{b0}$ was measured to be $(31110 \pm 50)$ Ohm.

The second correction factor to be considered is the Seebeck voltage generated at the junction between the gold platform and the Pt metal line. Indeed by using Pt to contact the gold platform we form a thermocouple which adds a series voltage $V_{Seeb}$ that depends on the temperature T of the membrane to the total bias $V_{BIAS}$. If a positive voltage bias is applied to the tip then $V_{Seeb}(T) = \Delta S \cdot (T - T_0)$ with $\Delta S = (S_{Pt} - S_{Au})$. The experimental characterization of this effect is a challenging task because it involves the disentanglement of the Seebeck effect from the temperature dependent $R_b(T)$. For Au-Pt wire thermocouples $\Delta S$ values of 7uV/K are reported[10]. The material quality can affect the value of the Seebeck coefficient even if it is hard to find a direct comparison in literature. Au-Pt[11] and Au-Cr[9] micro-thermocouples integrated in AFM tips are reported to have $\Delta S$ values that are very similar to the respective bulk values. Thus we use $\Delta S = 7.5$ µV/K with an estimated accuracy of 10%, which gives an offset $V_{Seeb} = 280$ µV at $\Delta T = 40$K.

## Error calculation

Possible sources of errors can be divided into systematic errors affecting all the data in the same manner, and random scatter represented for instance by the spread of points in the 2D histograms around the median (see Figure 3 in the main paper). The overall systematic error has been included as shaded region around the distribution median in Figure 4 of the main article.

The main sources of error in the measurements are the Seebeck offset $V_{Seeb}(T)$ and the uncertainty in the initial resistance value $R_{4P0}$ extracted from the *I-V* calibration. Errors on the underetched ratios, $R_b(T)$ and $I_{STM}$ are also considered even though they have a smaller influence.

Let *f* denote the measured quantity, in our case the median *f* of the 2D histogram. The absolute uncertainty $\delta f$ can be defined as function of the errors $\delta f_i$ induced by the single parameters *i* according to:

$$\delta f = \sqrt{\sum_{i=1}^{n} \delta f_i^{\,2}} \qquad (7)$$

where $\delta f_i = f_i(G_{\text{EL}}) - \bar{f}(G_{\text{EL}})$ represents the distance between the median $f_i$ and the reference $\bar{f}$ at every electrical conductance point $G_{\text{EL}}$. Following this approach we considered the extremes of the uncertainty range of the parameters listed above and we built the respective 2D histograms. From the medians $f_i$ of such distributions we calculated the errors $\delta f_i$ relative to the median $\bar{f}$ of the reference histogram. Finally, we used equation (7) to assess the total absolute error, $\delta f$. For instance, for the heater resistance at room temperature $R_{4P0} = (35.4 \pm 0.1)$ k$\Omega$ we used 35.4 k$\Omega$ as reference value and we calculated the 2D histograms with 35.5 k$\Omega$ and 35.3 k$\Omega$. The resulting medians define a positive and negative error for every $G_{\text{EL}}$ value. The same procedure is then applied for the other parameters. Note that the lower and upper limits of the uncertainty range are obtained from the sum of the negative and the positive errors respectively.

## 2D histograms

The 2D histogram shown in Figure 3 of the main article is built by counting the data-points that fall within a predefined interval of the electrical and the thermal conductance. For every trace, the reference value of $G_{\text{MEMS}}$ is calculated by selecting a part of the thermal conductance trace after breaking (below a threshold of 0.05 $G_0$) and by averaging over 10 data points. This allows to compensate for additional contributions to the thermal conductance stemming for instance from near field radiation or low electrically conducting hydrocarbons, which can be present on the gold surface. Some examples of single trace are shown in **Fig. S3**.

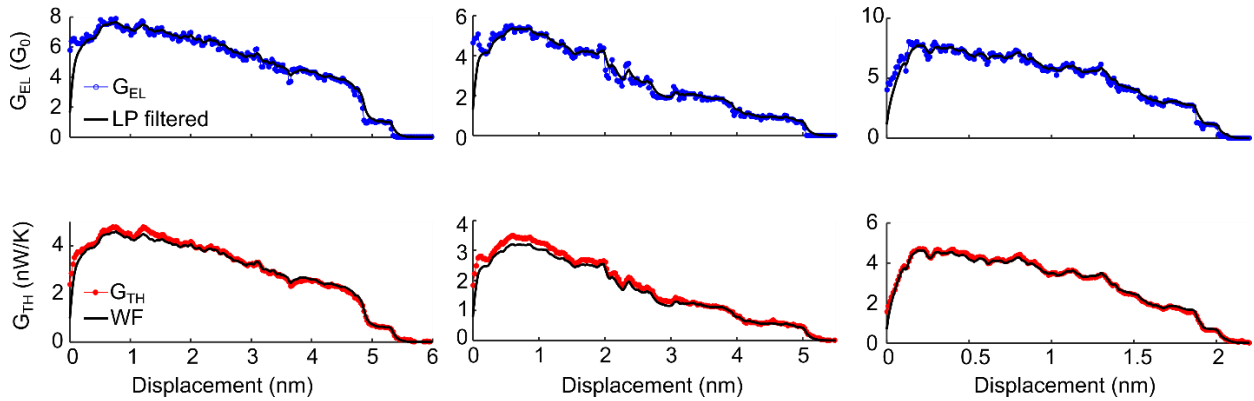

**Fig. S3| Example of single retraction traces taken during the measurement discussed in the main paper.** A voltage bias $V_{\text{BIAS}}$ = 50mV is applied to the STM tip and the average temperature difference is about 42K. For direct comparison the electrical conductance signal ($G_{\text{EL}}$) is first low-pass filtered (LP) with the thermal time constant of the MEMS ($\approx$ 30 ms) and then converted into thermal conductance ($G_{\text{TH}}$) by applying the Wiedemann-Franz (WF) law.

As discussed in the main paper, owing to near field radiation or spurious contaminants on the Au surfaces, parasitic heat paths can occur. These are manifested as non-constant thermal signals in the tunnelling region and beyond. In certain cases, such parasitic paths can even drastically reduce thermal sensitivity of our sensor and reduce the sensor temperature even without forming an electrical contact (thermal short cutting).

By using electrochemically etched tips and thermally cleaning the MEMS (see page 11) we were able to obtain data sets in which these effects are minimized and these background signals can be treated as a constant offset. For example, in Figure 2 of the main manuscript and in **Fig. S3** above, the thermal signal in the non-contact region (e.g. displacements larger than 5.5 nm, 5 nm and 2 nm, respectively) is used to define this constant offset included in the value of $G_{MEMS}$.

## Data Selection

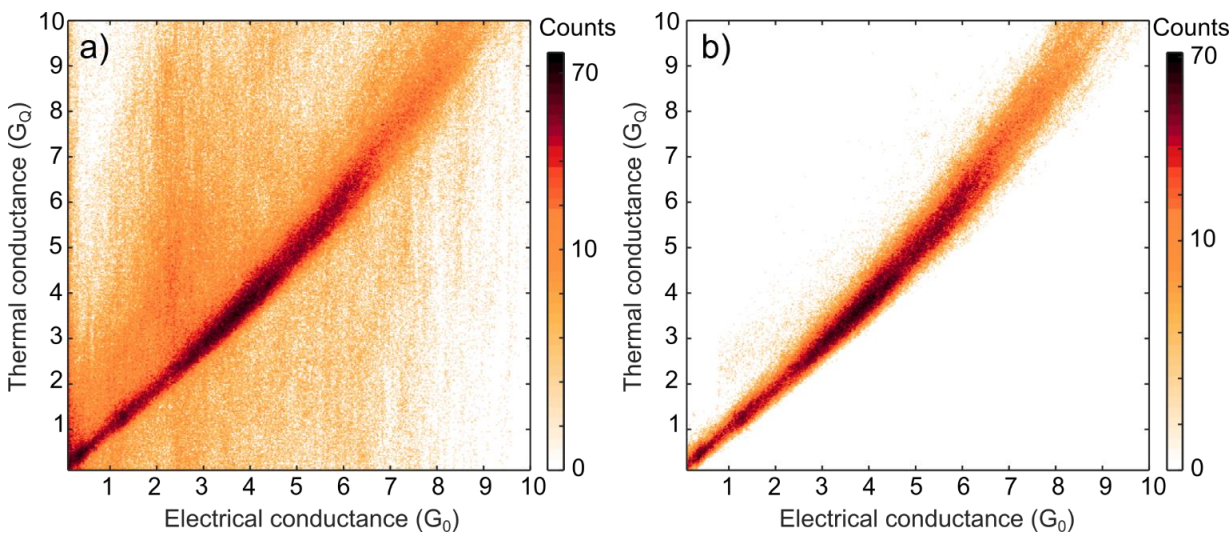

**Fig. S4| Effect of data selection on the 2D histogram shown in the main paper.** a) 2D histogram without trace selection obtained from 5000 traces. b) 2D Histogram obtained selecting the traces within the temperature range $\Delta T = (35\text{-}50)$ K corresponding to ~2000 traces.

Data selection in statistical evaluation of break junction experiments requires clear criteria to avoid arbitrary influencing the outcome. In these experiments, reconfiguration of tip and membrane surface can lead to thermal short cutting of the membrane and thereby loss of sensitivity. These situations are very clear in the data. It has three effects, of which either can be used as a criterion for data selection: first, the thermal short cutting is so large that the temperature of the MEMS sensor drops to below 10 degrees above ambient. Secondly, this may also leads to the occurrence of a significant slope in the thermal signal in the noncontact (tunnelling) region. Thirdly, the trace length between the predefined turn-around points (e.g. 5 $G_0$ and $10^{-5}$ $G_0$) can be unnaturally large. Typically, all three effects occur simultaneously.

The first criterion has been applied to the dataset in Figure 3. In **Fig. S4** we present, for comparison, the 2D-histograms before (**Fig. S4a**) and after (**Fig. S4b**) data selection, showing that the data omitted did not appear to have any correlated nature. Likewise, one of the data set for $V_{BIAS} = 300$ mV, which is reported in the table below, was reduced in this manner. The datasets shown in **Fig. S5** required no selection. All data shown was obtained with the same Au tip and MEMS. However, the reproducibility was further tested using several Au tips and MEMS devices. Within the respective uncertainties (currently about 30%) all these additional data support the conclusions drawn here.

## Control Datasets

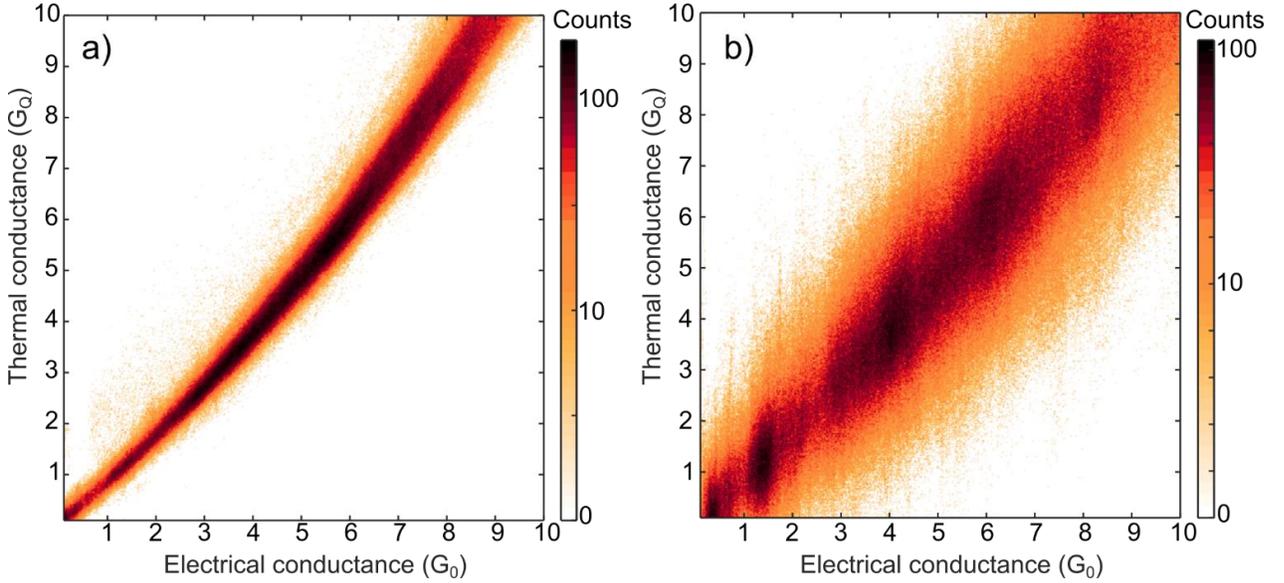

**Fig. S5| 2D histograms of thermal versus electrical conductance.** a) 2D histogram obtained at $V_{bias}$ = 50 mV. The same parameters as in the measurement reported in the main paper were used to show repeatability after several days. b) 2D histogram obtained with $V_{bias}$ = -300 mV. Here, Joule heating is strongly enhanced with respect to the 50 mV example. Both 2D histograms have been built from 5000 traces without trace selection with a linear binning and a bin size of 0.025 $G_Q$ and 0.025 $G_0$.

To provide further proof that the results shown in the main article are repeatable and do not depend on the voltage bias $V_{BIAS}$ applied, we show in **Fig. S5** 2D histograms measured with $V_{BIAS}$ = 50 mV (**Fig. S5a**) and $V_{BIAS}$ = -300 mV (**Fig. S5a**). At high bias (-300 mV) Joule dissipation in the junction and in the metallic line used to contact the Au platform leads to a net heating of the membrane until the thermal conductance of the contact is large enough to induce a net cooling. At low bias (50 mV) instead, net cooling of membrane upon contact is observed for any contact size. From the comparison of the measurements performed at different bias conditions it is clear that Joule dissipation alone cannot explain the observed correlation between thermal and electrical conductance.

Moreover, to summarize several control datasets, we quote here the average slope $G_{TH}/G_{EL}$ in the range of 1 to 5 $G_0$.

| $V_{BIAS}$ [mV] | 50 | 50 | 300 | 300 | -300 | -300 |
|---|---|---|---|---|---|---|
| $G_{TH}/(G_{EL}T)$ | 0.97 ± 0.15 (Figure 3) | 1.00 ± 0.15 **(Fig. S5a)** | 1.14 ± 0.29 | 0.88 ± 0.22 | 0.93 ± 0.23 | 0.97 ± 0.24 **(Fig. S5b)** |

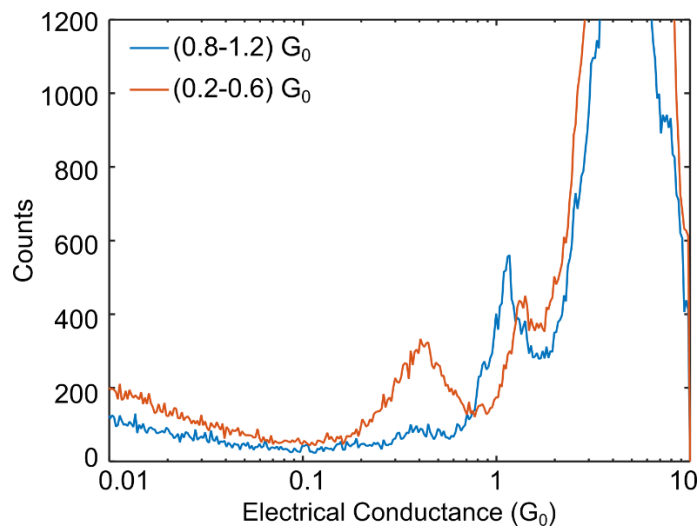

**Fig. S6 | Relation between the peaks at 1.3 $G_0$ and 0.4 $G_0$.** Electrical 1D conductance histograms built by selecting retraction traces showing at least 5 data-points plateau within the range indicated in the legend. Interestingly, the 1 $G_0$ peak shifts together with the intensity of the molecular peak (likely CO) at 0.4 $G_0$. This could mean that the molecule is attached to the gold electrodes prior to breaking, providing an additional path for charge transport. Both the orange and blue histograms were logarithmically binned and built with a similar number of traces, i.e. 695 and 472, respectively.

### Conductance peak shift

As discussed in the main paper, we expect that the contamination of the Au surface on the MEMS device with spurious carbon- and oxygen-containing molecules, such as carbon monoxide[3,12]. The discussed evidence for such a molecule can be the formation of a conductance peak below 1 $G_0$, and a shift or shoulder of the peak at 1 $G_0$ as observed elsewhere[13]. Therefore, two scenarios can be observed in individual opening traces: in one set of traces, no molecule is present, and a plateau at 1 $G_0$ is found. In a second set of traces, a molecule forms a parallel conductance channel at the contact (increasing the conductance of the plateau to above 1 $G_0$); the molecular junction may still be present even after breaking the Au-Au junction, leaving a conductance plateau below 1 $G_0$. It is possible to form histograms for the two subsets of data from Figure 3 of the main paper, as shown in **Fig. S6**. The fact that the histogram with peak at 1 $G_0$ (blue line) exhibit a drastically reduced number of counts in the region below 1 $G_0$ has taken as a strong evidence for these scenarios[13].

## Phonon contribution to thermal conductance

In Figure 4 of the main paper, we evaluated the phonon contribution to heat transport for different values of the electrical conductance. The details of the calculation are given in this section. Although the contribution of phonons falls within the experimental uncertainty, it is valuable to provide an estimate according to the theories and models available.

At temperatures greater than room temperature we do not expect the formation of long monoatomic chains when breaking the metallic contact at $G_{EL}=1$ $G_0$, but rather single atomic constrictions. In this limit we can describe heat transport within the 1D ballistic regime by using the Landauer-type approach for phonons[14,15], because both the thermal wavelength[16] and the phonon mean free path[17] are larger than the size of a single gold atom:

$$G_{\text{TH}} = \frac{k_B^2}{h} \sum_m \int_{x_m}^{\infty} dx \frac{x^2 e^x}{(e^x - 1)^2} \cdot T_m(x k_B T/\hbar) \qquad (8)$$

Here, the summation is over the phonon polarization modes $m$, $T_m$ are the transmission coefficients of the modes, $x_m = \omega_m(k=0)/k_B T$, assuming a small temperature difference $\Delta T$ across the junction, $k_B$ is the Boltzmann constant, $h$ the Planck constant, and $T$ the absolute temperature.

An upper bound for the thermal conductance $G_{\text{TH}}$ can be calculated with perfect adiabatic coupling of the thermal reservoirs to the ballistic contact, i.e. $T_m = 1$. To first approximation we can set the upper limit of the integral equal to the Debye frequency of gold $\omega_D = k_B T_D/\hbar$ with $T_D = 165$ K and take into account the 3 acoustic modes of the junction[18]. With these approximations equation (8) reduces to:

$$G_{\text{TH}} = 3 \frac{k_B^2}{h} \int_0^{x_D} dx \frac{x^2 e^x}{(e^x - 1)^2} \qquad (9)$$

where $x_D = \hbar \omega_D / k_B T$. At an average temperature $T = 320$ K we find that $G_{\text{TH,ph}} \cong 0.22$ $G_{\text{TH,el}}$ at $G_{EL} = 1$ $G_0$. However, according to the microscopic details of the junction, in particular to the shape of the electrodes, a decrease of $G_{\text{TH,ph}}$ by about an order of magnitude with respect to the case of an ideal geometry is expected[19]. Therefore, we can assume a minimum contribution of phonons to heat transport of $G_{\text{TH,ph}} \cong 0.02$ $G_{\text{TH,el}}$.

To extend this uncertainty range to higher values of electrical conductance, we can consider that the number of both the electron and the phonon channels involved in transport should both increase in proportion to the area of the contact, $A$, at least for sufficiently large contacts. Then the ratio $G_{\text{TH,ph}}/G_{\text{TH,el}}$ will be constant with the electrical conductance. For contacts of very few conductance channels this does not hold strictly, because the correspondence of the cross sectional area of one atom to one conduction channel is valid only for the single atom case[20]. Nevertheless, the range $G_{\text{TH,ph}} = (0.02 \div 0.22)G_{\text{TH,el}}$ from the above considerations should account for these uncertainties and approximations. To find a medium value for the phonon contribution to the thermal conductance (red line in Figure 4 of the main article) we can modify equation (9) as

$$G_{\text{TH}} = 3N \frac{A_{\text{EL}}}{A_{\text{PH}}} \frac{k_B^2}{h} \int_0^{x_D} dx \frac{x^2 e^x}{(e^x - 1)^2} \qquad (10)$$

where $N$ is the number of electron channels available for conduction, and, $A_{EL}$ and $A_{PH}$ denote the area occupied by an electron and phonon channel, respectively. It is possible to estimate $A_{EL} = \lambda_F^2/\pi$ and $A_{PH} = \lambda_{TH}^2/\pi$ where $\lambda_F$ is the electron Fermi wavelength and $\lambda_{TH}$ the dominant phonon wavelength when considering thermal conductance[21]. For gold $\lambda_F \cong 0.52$ nm and $\lambda_{TH} \cong 2\lambda_0 \cong 0.83$ nm where $\lambda_0$ represents the smallest allowed phonon wavelength as indicated by Dames[16].

Finally, we can obtain similar results if we apply kinetic theory to treat ballistic heat transport through point contacts[22], i.e. by extending a classical theory to the range of our experiments. To estimate $A$ we use Sharvin's formula for electron transport $G_S = G_0(\pi A/\lambda_F^2 - P/2\lambda_F)$ as shown by Torres et al.[23], where $A = \pi R^2, P = 2\pi R$, with $R$ the contact radius. By calculating $R$ we can then apply Wexler's formula in the ballistic limit to describe phonon transmission: $G_{TH} = 3\pi R^2 k/4\,\Lambda$, where k is the bulk phonon thermal conductivity and $\Lambda$ the phonon mean free path. From Ref.[17] we estimate for gold $\Lambda \cong 2$ nm and $k \cong 2$ W/mK obtaining $G_{TH,ph} \cong 0.14 G_{TH,el}$ at $G_{EL} = 8\,G_0$, in good agreement with the range calculated above.